\documentclass[aps,preprint,onecolumn]{revtex4}%
\usepackage{amssymb}
\usepackage{amsmath}
\usepackage{amsfonts}
\usepackage{graphicx}%
\providecommand{\U}[1]{\protect\rule{.1in}{.1in}}

\begin{document}
\title{Relativistic ideal Bose gas in Harmonic traps}
\author{Xiang-Mu Kong}
\thanks{Corresponding author}
\email{kongxm@mail.qfnu.edu.cn (X.-M.Kong)}
\author{Ying Wang}
\author{Cong-Fei Du}
\affiliation{College of Physics and Engineering, Qufu Normal university, Qufu 273165, China }

\begin{abstract}
Using semiclassical approximation method, Bose-Einstein condensation (BEC) of
a relativistic ideal boson gas (RIBG) with and without antibosons in
three-dimensional (3-D) harmonic traps is investigated. The BEC transition
temperature $T_{c}$ and the Helmholtz free energy at $T_{c}$ are calculated.
The effect of the rest mass of the boson on the properties of the system is
also studied. We find that $T_{c}$ of RIBG is higher than that of the
nonrelativistic approximation. The RIBG with antibosons is also investigated
and it is found that the Helmholtz free energy of the system with antibosons
at $T_{c}$ is lower than that of the system without antibosons. It implies
that the system with antibosons is more stable.

\end{abstract}
\volumeyear{year}
\volumenumber{number}
\issuenumber{number}
\eid{identifier}
\startpage{1}
\endpage{ }
\maketitle

\section{ Introduction}

Bose-Einstein condensation (BEC) has attracted a great deal of attention since
its theoretical prediction by Einstein in 1925 \cite{Einstein1924}. Initially,
in the theoretical research on the phenomenon, the relativistic corrections
are neglected near the BEC transition temperature $T_{c}$ \cite{Huang1987,C.
J. Pethick2002,F. Dalfovo1999}. Lately, it is found that in the universe some
systems consist of bosons of\ very small rest mass. For example, the rest mass
of a pair of neutrinos is about $10^{^{-30}}$g in the neutrinos system
\cite{P. T. Landsberg1965,Su. G.2006}. For such boson systems, the
relativistic effects should be considerable. So particular attention has been
given to the behavior of the relativistic boson gas.

The properties of the relativistic ideal boson gas (RIBG) have been discussed
in early papers \cite{May1964,P. T.
Landsberg1965,Landsberg1981,Beckmann1979,Beckmann1982}. In 1965, Landsberg and
Dunning-Davies studied the critical temperature $T_{c}$ and the specific-heat
anomaly at $T_{c}$ of the RIBG \cite{P. T. Landsberg1965,Landsberg1981}. On
the basis of the above works, Beckmann \textit{et al. }investigated the
behavior of RIBG with and without mass in all space dimensions (d)
\cite{Beckmann1982}. It was found that for the massless gas, the specific-heat
has a gap in $d>2$, while for the massive gas, there was a gap in $d>4$.
Although the research on the BEC in the relativistic system has a long
history, a complete treatment using the quantum field theory has not been
carried out until 1980s, in which the pair creation of boson-antiboson is
considered \cite{Haber11981,Haber1982,Singh 1983,Singh1984,Frota1989}. Haber
and Weldon firstly gave the high-temperature expansions for thermodynamic
functions of the RIBG by taking into account antibosons \cite{Haber11981}.
Frota and Singh \textit{et al}. exhibited in detail the thermodynamic
properties of d-dimensional system with antibosons \cite{Frota1989,Singh
1983,Singh1984}. Moreover, the more exact treatments of BEC in RIBG are given
\cite{Grether 2007}. Recently, a relativistic BCS-BEC crossover at zero
temperature and finite temperature has been studied by the quantum field
theory \cite{He 2007,He1 2007,Nishida 2005}.

It is worth noticing that the previous investigations on the properties of the
relativistic boson gas were mainly focused on the systems in the absence of
external potential. As we know, the external potential is very important to
control the characteristics of the Bose gas \cite{Bagnato 1987,Salasnich
2000}. Su \textit{et al.} discussed the BEC of RIBG in a general external
potential by using classical Hamiltonian \cite{Su. G.2006}. In this paper, we
study the behavior of the RIBG with and without antibosons in 3-D harmonic
traps using the quantum theory.

The outline of this paper is as follows. In section \ref{without}, we study
the BEC transition temperature and Helmholtz free energy at $T_{c}$.
Meanwhile, we also investigate the effects of the different values of mass on
the BEC of the RIBG. In Sec. \ref{with}, we introduce antibosons and discuss
the BEC in such a system with antibosons, and Sec. \ref{summary} is a brief summary.

\section{Relativistic Bose-Einstein condensation\label{without}}

Consider a RIBG system of $N$ bosons in 3-D harmonic traps, \ and the
potential function $V(\mathbf{r})$ can be expressed as%

\begin{equation}
V(\mathbf{r})=\frac{m}{2}\left(  \omega_{x}^{2}x^{2}+\omega_{y}^{2}%
y^{2}+\omega_{z}^{2}z^{2}\right)  , \label{V}%
\end{equation}
where $m$ and $\mathbf{r}$ are the rest mass and coordinate of each boson,
respectively. $\omega_{x}$, $\omega_{y}$ and $\omega_{z}$ are the frequencies
of anisotropy harmonic traps. By means of the standard
Rayleigh-Schr\"{o}dinger perturbation \cite{Cohen 1986,Harvey1972}, the total
energy of each boson $E_{n_{x}\text{,}n_{y}\text{,}n_{z}}$ can be given by%
\begin{equation}
E_{n_{x}\text{,}n_{y}\text{,}n_{z}}=E^{(0)}+E^{(1)}, \label{E}%
\end{equation}
in which%

\begin{align}
E^{(0)}  &  =3mc^{2}+\hbar\left[  n_{x}\omega_{x}+n_{y}\omega_{y}+n_{z}%
\omega_{z}+\frac{1}{2}\left(  \omega_{x}+\omega_{y}+\omega_{z}\right)
\right]  ,\label{E0}\\
E^{(1)}  &  =\frac{3\hbar^{2}}{16mc^{2}}\left[  n_{x}^{2}\omega_{x}^{2}%
+n_{y}^{2}\omega_{y}^{2}+n_{z}^{2}\omega_{z}^{2}+n_{x}\omega_{x}^{2}%
+n_{y}\omega_{y}^{2}+n_{z}\omega_{z}^{2}+\frac{1}{2}\left(  \omega_{x}%
^{2}+\omega_{y}^{2}+\omega_{z}^{2}\right)  \right]  ,
\end{align}
where $\hbar$ is the Planck constant, and $c$ is the speed of light. $n_{x}$,
$n_{y}$ and $n_{z}$ are 3-D quantum numbers of the boson energy.

To describe RIBG in grand canonical ensemble, the usual expression for the
number of particles $N$ in statistical mechanics is%

\begin{equation}
N=\sum_{n}\rho_{n}=\sum_{n_{x},n_{y,}n_{z}}\frac{1}{e^{\beta\left(
E_{n_{x}\text{,}n_{y}\text{,}n_{z}}-\mu\right)  }-1}, \label{N}%
\end{equation}
where $\beta=1/k_{B}T$, $k_{B}$ is the Boltzmann constant, and $\mu$ is the
chemical potential. $\rho_{n}$ is the average number of bosons in the state of
energy $E_{n_{x}\text{,}n_{y}\text{,}n_{z}}$. The number of particles in the
ground state $N_{0}$ becomes to take on macroscopic values corresponding to
the onset of BEC. It is convenient to separate out the lowest energy
$E_{0\text{,}0\text{,}0}$ from the sum Eq. (\ref{N}), so one can write%

\begin{equation}
N-N_{0}=\sum_{n_{x},n_{y,}n_{z}\neq0}\frac{1}{\exp[\beta\left(  E^{(0)}%
+E^{(1)}-\mu\right)  ]-1}, \label{N-N0}%
\end{equation}
when BEC occurs, the chemical potential equals to the energy of the lowest
state $E_{0\text{,}0\text{,}0}$, i.e., $\mu_{c}\rightarrow3mc^{2}+\hbar\left(
\omega_{x}+\omega_{y}+\omega_{z}\right)  /2+3\hbar^{2}\left(  \omega_{x}%
^{2}+\omega_{y}^{2}+\omega_{z}^{2}\right)  /32mc^{2}$. Eq. (\ref{N-N0}) can be
reduced as%

\begin{equation}
N-N_{0}=\sum_{n_{x},n_{y,}n_{z}\neq0}\frac{1}{\exp[\beta(\varepsilon
^{(0)}+\varepsilon^{(1)})]-1}, \label{n-n0}%
\end{equation}
where%

\begin{align}
\varepsilon^{(0)}  &  =\hbar(n_{x}\omega_{x}+n_{y}\omega_{y}+n_{z}\omega
_{z}),\label{e0}\\
\varepsilon^{(1)}  &  =\frac{3\hbar^{2}}{16mc^{2}}(n_{x}^{2}\omega_{x}%
^{2}+n_{y}^{2}\omega_{y}^{2}+n_{z}^{2}\omega_{z}^{2}+n_{x}\omega_{x}^{2}%
+n_{y}\omega_{y}^{2}+n_{z}\omega_{z}^{2}).
\end{align}

In order to evaluate the above sum explicitly, it is assumed that the energy
spacing becomes smaller when $N\rightarrow\infty$. So sums (\ref{n-n0}) become
integrals as%

\begin{equation}
N-N_{0}=\int_{0}^{\infty}\frac{dn_{x}dn_{y}dn_{z}}{\exp[\beta(\varepsilon
^{(0)}+\varepsilon^{(1)})]-1}. \label{n-n0f}%
\end{equation}

This approximate transformation is good when trapped particles become very
large and $k_{B}T\gg\hbar\omega_{x},\hbar\omega_{y},\hbar\omega_{z}.$ So the
BEC transition temperature of the RIBG can be obtained by taking
$N_{0}\rightarrow0$, one gets%

\begin{equation}
N=\int_{0}^{\infty}\frac{dn_{x}dn_{y}dn_{z}}{\exp[\beta_{c}(\varepsilon
^{(0)}+\varepsilon^{(1)})]-1}, \label{nf}%
\end{equation}
where $\beta_{c}=1/k_{B}T_{c}$. Fig. 1 displays the exact $T_{c}$ numerically
versus $N$ for different values of rest mass. It is found that the BEC
transition temperature increases with the decrease of the rest mass of the
particle, which is similar to the case of a uniform Bose gas \cite{P. T.
Landsberg1965}. It is explained that the relativistic effect is obvious for
the boson system with very small rest mass. The transition temperature is
compared with the result based on the nonrelativistic approximation (see Fig.
2). It is shown that the relativistic effect results in the increasing of the
BEC transition temperature, which disagrees with the existing result \cite{Su.
G.2006}.

The free energy at the critical point can be also investigated. We now discuss
the Helmholtz free energy at $T_{c}$ in harmonic traps. According to the
partition function of the system, the Helmholtz free energy can be expressed as%

\begin{equation}
F=\mu_{c}N+k_{B}T_{C}\int_{0}^{\infty}dn_{x}dn_{y}dn_{z}\{\ln[1-\exp\left[
-\beta_{c}(E_{n_{x}\text{,}n_{y}\text{,}n_{z}}-\mu_{c})\right]  ]\},
\label{fb}%
\end{equation}
where $\beta_{c}$ is obtained numerically for a given value of $N$ and
$E_{n_{x}\text{,}n_{y}\text{,}n_{z}}$ is given by Eq. (\ref{E}). Fig. 3 shows
the Helmholtz free energy at $T_{c}$ as a function of $N$ for different values
of the rest mass. It can be seen that the smaller the rest mass of boson is,
the lower the Helmholtz free energy is, namely, the system is more stable.

\section{Antiboson and Bose-Einstein condensation\label{with}}

To our knowledge, at sufficiently high temperature the quantum field theory
requires consideration of particle-antiparticle pair production. If
$\overline{N}$ is the number of antibosons, the system is governed by the
conservation of the number $Q=N-\overline{N}$, rather than of the numbers $N$
and $\overline{N}$ separately. According to Ref \cite{Harvey1972}, $E=\pm
E_{n}.$ For bosons we have $E=+E_{n}$ and for antibosons we have $E=-E_{n}.$
So $N=\sum_{n}\rho_{n}$ is replaced by%

\begin{align}
Q  &  =N-\overline{N}=\sum_{n}\left(  \rho_{n}-\overline{\rho}_{n}\right)
\label{N=N}\\
&  =\sum_{n_{x},n_{y,}n_{z}}\left[  \frac{1}{e^{\beta\left(  E_{n_{x}%
\text{,}n_{y}\text{,}n_{z}}-\mu\right)  }-1}-\frac{1}{e^{\beta\left(
E_{n_{x}\text{,}n_{y}\text{,}n_{z}}+\mu\right)  }-1}\right]  ,\nonumber
\end{align}
where $\overline{\rho}_{n}$ is the average number of antibosons in the state
of energy $-E_{n_{x}\text{,}n_{y}\text{,}n_{z}}$. Since the number of bosons
(antibosons) in various states must be positively defined, we have $|\mu|$
$\leq\mu_{c}$. Using the similar method as in Sec. \ref{without}, the BEC
critical temperature can be obtained by%

\begin{align}
&  N-\overline{N}\label{N=NF}\\
&  =\int_{0}^{\infty}dn_{x}dn_{y}dn_{z}\left[  \frac{1}{\exp[\beta
_{c}(\varepsilon^{(0)}+\varepsilon^{(1)})]-1}-\frac{1}{\exp[\beta
_{c}(\varepsilon^{(0)}+\varepsilon^{(1)}+2\mu_{c})]-1}\right]  .\nonumber
\end{align}
Fig. 4 gives the behavior of $T_{c}$ numerically obtained from Eq.
(\ref{N=NF}) for the different values of the rest mass. It can be seen that
the behavior of $T_{c}$ is similar to the case of the system without
antibosons. However, comparing to the result of the system without antibosons,
we find that $T_{c}$ of the system with antibosons is higher (see Fig. 5).
This means as the decreasing of the temperature, BEC of the system considering
antibosons occurs firstly. It implies that the system with antibosons is more
stable, i.e., has a lower Helmholtz free energy at $T_{c}$. This will be shown
to be the case indeed.

The Helmholtz free energy of the system with antibosons can be also given by%

\begin{align}
F  &  =\mu_{c}(N-\overline{N})\label{fbb}\\
&  +k_{B}T_{C}\int_{0}^{\infty}dn_{x}dn_{y}dn_{z}\{\ln[1-\exp\left[
-\beta_{c}(E_{n_{x}\text{,}n_{y}\text{,}n_{z}}-\mu_{c})\right]  ]+\ln
[1-\exp\left[  -\beta_{c}(E_{n_{x}\text{,}n_{y}\text{,}n_{z}}+\mu_{c})\right]
]\},\nonumber
\end{align}
where $F$ is just Eq. (\ref{fb}) but with the second log term caused by
antibosons. In Fig. 6, we find that for a given value of $N$ the smaller the
rest mass is, the more stable the system is. However, it is found that $F$ of
the system considering antibosons is always lower than that of the system
without antibosons, which proves the above speculation (see Fig. 7).

\section{Summary\label{summary}}

In this paper, we have studied the BEC of the relativistic Bose gas with and
without antibosons in harmonic traps by the quantum theory. It can be seen
that the BEC transition temperature increases with the decrease of the rest
mass of the particle. By using the quantum energy spectrum, we find the
relativistic BEC transition temperature is higher than that of the
nonrelativistic approximation. $T_{c}$ of the system with antibosons is higher
than that of the system without antibosons, and the calculation of the
Helmholtz free energy at $T_{c}$ further reveals that the system with
antibosons is more stable.

\section{Acknowledgment}

This work is supported by the National Natural Science Foundation of China
under Grant No. 10775088, the Shandong Natural Science Foundation under Grant
No. Y2006A05 and the Science Foundation of Qufu Normal University.

\newpage

\bigskip Figure Captions

Fig. 1. The BEC transition temperatures $T_{c}s$ (in units of $K$) versus the
number of particles $N$ without antibosons for different mass and the harmonic
frequencies $\omega_{x}=\omega_{y}=10^{17}$ $Hz,\omega_{z}=2\ast10^{17}$ $Hz$.
The solid, dotted and dashed lines represent the various values of mass
$m=3\ast10^{-34}$ $kg,$ $3\ast10^{-33}$ $kg$ and $3\ast10^{-32}$ $kg,$ respectively.

Fig. 2. Difference between the BEC transition temperatures $T_{c}s$ (in units
of $K$) for relativistic and nonrelativistic approximation with the mass $m=$
$3\ast10^{-32}$ $kg,$ harmonic frequencies $\omega_{x}=\omega_{y}=10^{17}$
$Hz,\omega_{z}=2\ast10^{17}$ $Hz.$ The solid and dotted lines stand for the
relationship between the transition temperatures $T_{c}s$ and the number of
particles $N$ for nonrelativistic and relativistic cases, respectively.

Fig. 3. The Helmholtz free energy $F$ (in units of $\mu N$) versus the number
of particles $N$ without antibosons at $T_{c}$ for different mass and the
harmonic frequencies $\omega_{x}=\omega_{y}=10^{17}$ $Hz,\omega_{z}%
=2\ast10^{17}$ $Hz$. The solid, dotted and dashed lines represent the various
values of mass $m=3\ast10^{-32}$ $kg,$ $1\ast10^{-32}$ $kg$ and $7\ast
10^{-33}$ $kg,$ respectively.

Fig. 4. The BEC transition temperatures $T_{c}s$ (in units of $K$) versus the
number of particles $N$ with antibosons for different mass and the harmonic
frequencies $\omega_{x}=\omega_{y}=10^{17}$ $Hz,\omega_{z}=2\ast10^{17}$ $Hz$.
The solid, dotted and dashed lines represent the various values of mass
$m=3\ast10^{-34}$ $kg,$ $3\ast10^{-33}$ $kg$ and $3\ast10^{-32}$ $kg,$ respectively.

Fig. 5. Difference between the BEC transition temperatures $T_{c}s$ (in units
of $K$) with and without antibosons for the mass $m=$ $3\ast10^{-32}$ $kg,$
harmonic frequencies $\omega_{x}=\omega_{y}=10^{17}$ $Hz,\omega_{z}%
=2\ast10^{17}$ $Hz.$ The solid and dotted lines stand for the relationship
between the transition temperatures $T_{c}s$ and the number of particles $N$
for the cases without and with antibosons, respectively.

Fig. 6. The Helmholtz free energy $F$ (in units of $\mu N$) versus the number
of particles $N$ with antibosons at $T_{c}$ for different mass and the
harmonic frequencies $\omega_{x}=\omega_{y}=10^{17}$ $Hz,\omega_{z}%
=2\ast10^{17}$ $Hz$. The solid, dotted and dashed lines represent the various
values of mass $m=3\ast10^{-32}$ $kg,$ $1\ast10^{-32}$ $kg$ and $7\ast
10^{-33}$ $kg,$ respectively.

Fig. 7. Difference between the Helmholtz free energy $F$ (in units of $\mu N$)
with and without antibosons at $T_{c}$ for the mass $m=$ $3\ast10^{32}$ $kg,$
harmonic frequencies $\omega_{x}=\omega_{y}=10^{17}$ $Hz,\omega_{z}%
=2\ast10^{17}$ $Hz.$ The solid and dotted lines stand for the relationship
between the Helmholtz free energies $F$ and the number of particles $N$ for
the cases without and with antibosons, respectively.

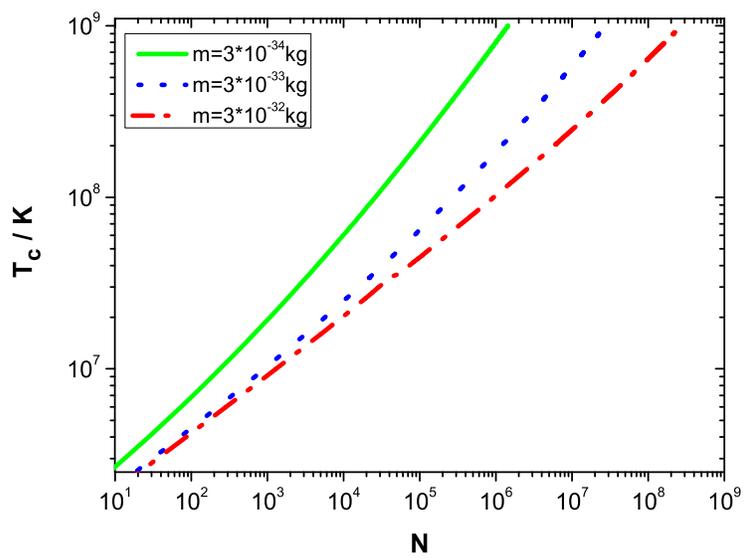
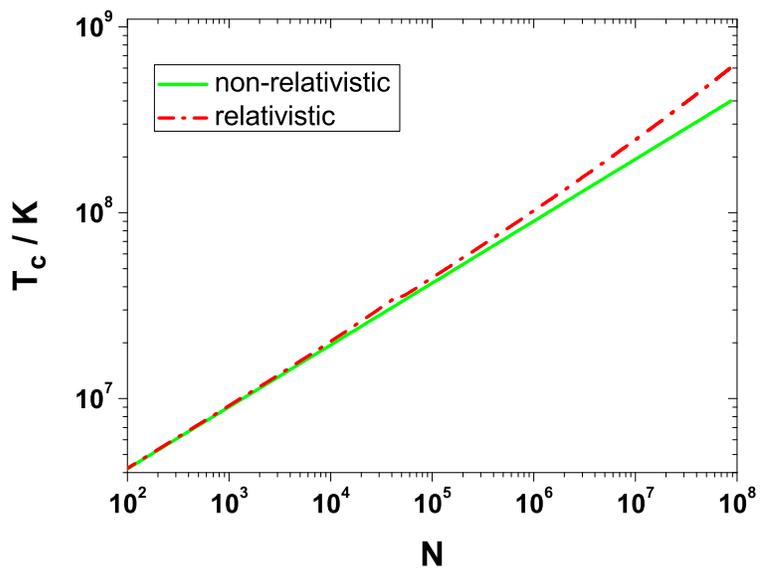
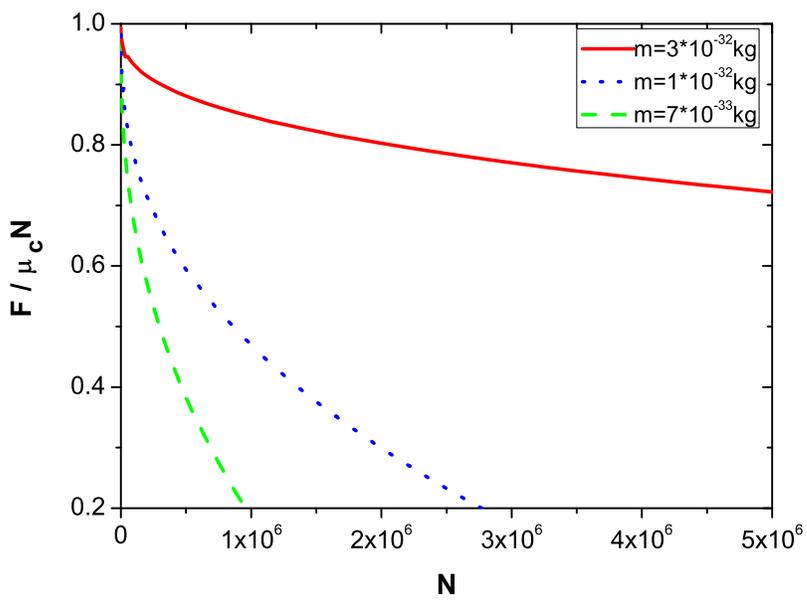
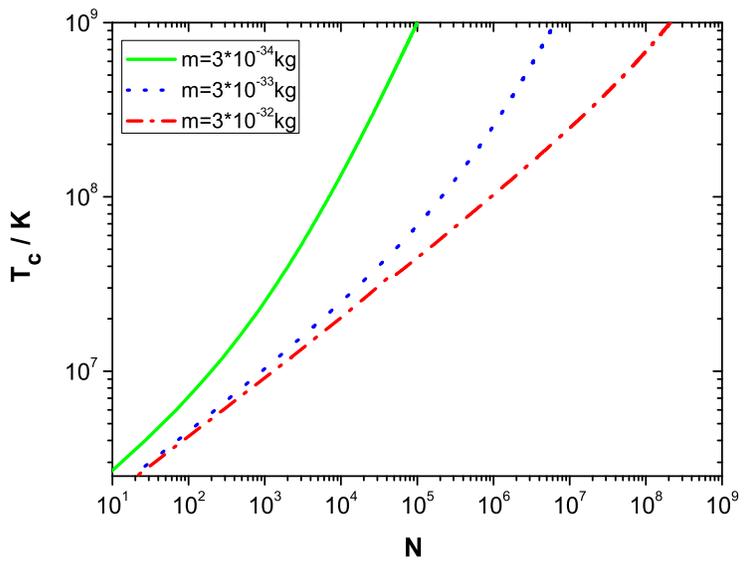
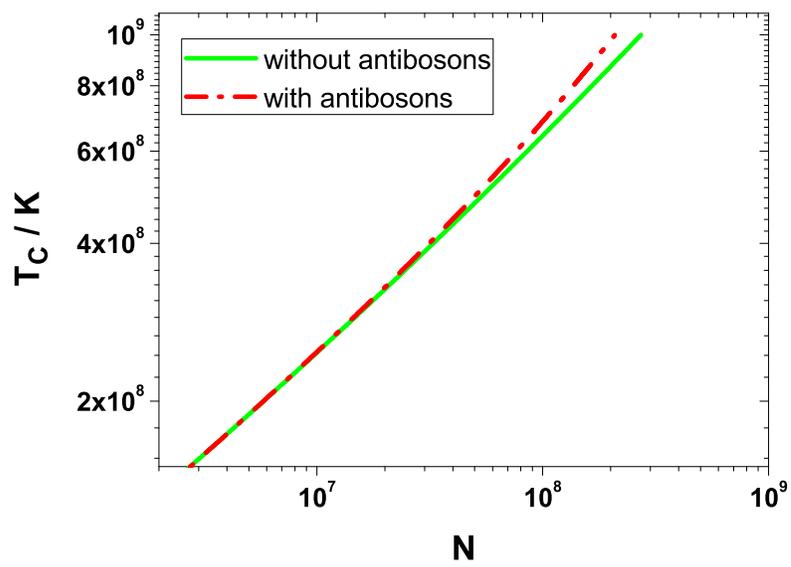
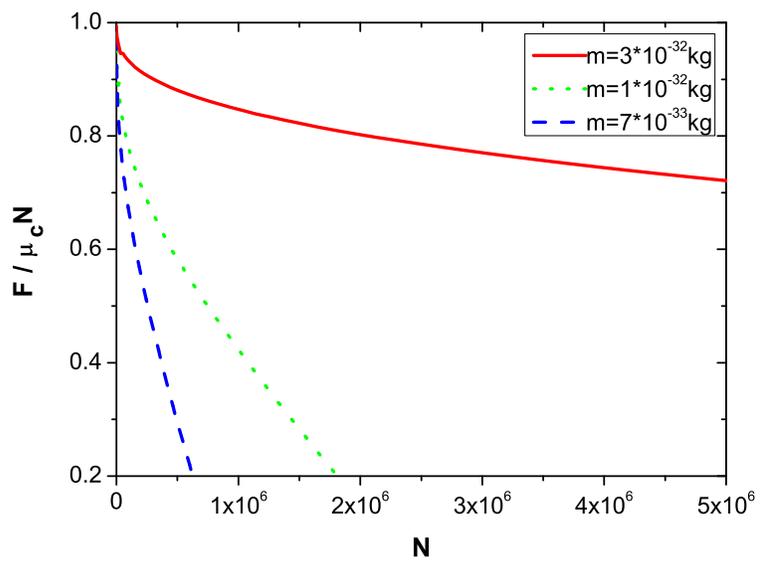
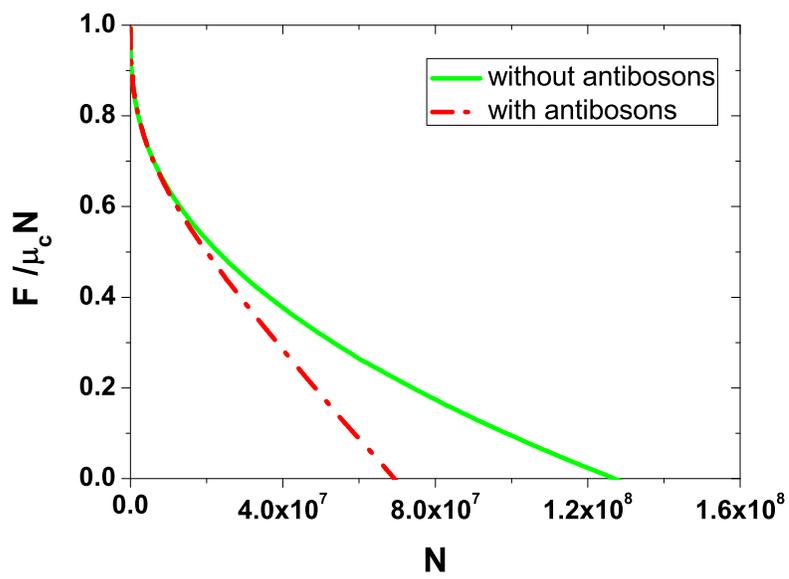
\end{document}